\documentclass[conference]{IEEEtran}
\IEEEoverridecommandlockouts
\usepackage{cite}
\usepackage{amsmath,amssymb,amsfonts}
\usepackage[noend]{algpseudocode}
\usepackage[ruled]{algorithm}
\usepackage{graphicx}
\usepackage{subcaption}

\algnewcommand\algorithmicforeach{\textbf{for each}}
\algdef{S}[FOR]{ForEach}[1]{\algorithmicforeach\ #1\ \algorithmicdo}

\algnewcommand{\Inputs}[1]{%
	\State \textbf{Inputs:}
	\Statex \hspace*{\algorithmicindent}\parbox[t]{\linewidth}{\raggedright #1}
}

\algnewcommand{\Initialize}[1]{%
	\State \textbf{Initialize:}
	\Statex \hspace*{\algorithmicindent}\parbox[t]{\linewidth}{\raggedright #1}
}

\algnewcommand{\Output}[1]{%
	\State \textbf{Output:}
	\Statex \hspace*{\algorithmicindent}\parbox[t]{\linewidth}{\raggedright #1}
}
\algnewcommand{\Definitions}[1]{%
	\State \textbf{Definitions:}
	\Statex \hspace*{\algorithmicindent}\parbox[t]{.95\linewidth}{\raggedright #1}
}

\algrenewcommand\textproc{}

\makeatletter
\newcommand{\linebreakand}{%
  \end{@IEEEauthorhalign}
  \hfill\mbox{}\par
  \mbox{}\hfill\begin{@IEEEauthorhalign}
}
\makeatother

\usepackage{graphicx}
\usepackage{textcomp}
\usepackage{xcolor}
\def\BibTeX{{\rm B\kern-.05em{\sc i\kern-.025em b}\kern-.08em
    T\kern-.1667em\lower.7ex\hbox{E}\kern-.125emX}}

\usepackage[T1]{fontenc}
\usepackage{comment}

\begin{document}

\title{Aero-LLM: A Distributed Framework for Secure UAV Communication and Intelligent Decision-Making}
\author{
    \IEEEauthorblockN{
        Balakrishnan Dharmalingam, Rajdeep Mukherjee, Brett Piggott, Guohuan Feng, Anyi Liu
    }
    \IEEEauthorblockA{Department of Computer Science and Engineering\\
        Oakland University\\
        Rochester, Michigan, U.S.A. \\
        \{bdharmalingam, rajdeepmukherje, bapiggott, gfeng, anyiliu\}@oakland.edu}
}    

\maketitle

\begin{abstract}
Increased utilization of unmanned aerial vehicles (UAVs) in critical operations necessitates secure and reliable communication with Ground Control Stations (GCS). This paper introduces Aero-LLM, a framework integrating multiple Large Language Models (LLMs) to enhance UAV mission security and operational efficiency. Unlike conventional singular LLMs, Aero-LLM leverages multiple specialized LLMs for various tasks, such as inferencing, anomaly detection, and forecasting, deployed across onboard systems, edge, and cloud servers. This dynamic, distributed architecture reduces performance bottleneck and increases security capabilities. Aero-LLM’s evaluation demonstrates outstanding task-specific metrics and robust defense against cyber threats, significantly enhancing UAV decision-making and operational capabilities and security resilience against cyber attacks, setting a new standard for secure, intelligent UAV operations.

\end{abstract}


\section{Introduction}\label{section:introduction}
 With the increasing deployment of unmanned aerial vehicles (UAVs) in mission-critical operations, securing the communication channels between UAVs and Ground Control Stations (GCS) becomes paramount. The integrity and confidentiality of the transmitted data must be ensured. In the evolving landscape of artificial intelligence, integrating Large Language Models (LLMs) with system and software security transforms defense mechanisms against cyber threats. LLMs, with their advanced capabilities in language comprehension and generation, offer significant potential to enhance detection and defensive strategies against cyber adversaries~\cite{wang2023chatgpt, 10177704, DBLP:journals/corr/abs-2202-01142, Jigsaw, ferrag2023revolutionizing, sandoval2023lost}. The data exchange between UAVs and GCS, particularly sensor data, involves time series with varying sampling rates and occasional data gaps, necessitating meticulous preprocessing to ensure the data's usability for fine-tuning LLMs.

In this paper, we present Aero-LLM, a novel framework that integrates different types of LLMs as a team for collaboration and information sharing for UAV flying missions. Specifically, different types of LLMs show capabilities in inferencing, anomaly detection, and forecasting. The LLMs team is strategically placed onboard, at the edge, or in the cloud to provide a robust and effective defense against various cyber exploits. Compared with the state-of-the-art (SOTA) foundation LLMs, such as Llama~\cite{Llama2}, Gemini~\cite{Gemini}, Mistral~\cite{Mistral}, and DBRX~\cite{DBRX}, Aero-LLM is not designed to outperform their benchmark capabilities. Instead, it fine-tuned specialized LLMs, which focus on particular tasks and data. The striking feature of Aero-LLM is two-fold. First, it leverages a team of LLMs, whose size might be considerably smaller than the versatile but all-in-one LLMs. Each LLM takes the cross-sectional or time-series data and accomplishes specific tasks accordingly. The distributed architecture also potentially introduces intelligent agents into the scene, which orchestrates and moderates the coordination among LLMs.
Second, it allows the partial LLMs to be offloaded to the computing units on edge servers or cloud servers. Thus, some inference tasks can be performed by more powerful GPUs off-board. To train special-skilled LLMs, we collectively applied two fine-tuning technologies: 1) \textit{supervised fine-tuning} (SFT); and 2) \textit{reinforcement learning from human feedback} (RLHF). To evaluate the performance of Aero-LLM, we tested two types of LLM on different tasks. Specifically, we use small-scale LLMs (OPT-350m and OPT-125m\cite{zhang2022opt}) to mimic the UAV in network communication with the GCS. We use time-sensitive LLMs  (e.g., TimesNet) to detect anomalies and forecast future activities. Our evaluation results demonstrate that Aero-LLM is able to integrate the generative capabilities of various LLMs with high accuracy, in terms of accuracy, precision, recall, F1 score, and low error rates. Moreover, the fine-turned LLMs demonstrate low memory footprints in terms of VRAM usage. The contributions of this paper are summarized as follows:

\begin{itemize}
    \item We designed and implemented Aero-LLM, a novel framework by assembling diverse LLMs for collaborative functioning, optimizing UAV missions through task specialization. This method advances beyond traditional large-scale LLMs by emphasizing precise, mission-critical capabilities enhanced through supervised fine-tuning and reinforcement learning from human feedback.
    \item The Aero-LLM promotes a scalable, distributed architecture that efficiently utilizes on-board, edge, and cloud computing resources, which ensures that Aero-LLM maintains high performance and cybersecurity standards across varied operational environments and computational capacities.
    \item We tested Aero-LLM's capabilities and observed superior task-specific performance, such as high accuracy, precision, recall, and F1 scores with minimal error rates. Moreover, its distributed nature offers robust protection against cyber threats, highlighting an advanced cybersecurity model within UAV operations.
    \end{itemize}

We organize the paper as follows: Section~\ref{section:related-work} researches the related work in the field. Section~\ref{section:threat-model} briefly describes the capabilities of an adversary and the attacking scenario. Section~\ref{section:system-design} provides the detailed steps of constructing Aero-LLM. Section~\ref{section:evaluation} presents the experimental results. Section~\ref{section:conclusion} concludes the paper and suggests our future directions.

\section{Related Work} \label{section:related-work}
This section briefly reviews the related work in three research domains: 1) LLM application in IoT and embedded systems; 2) Smaller-scale LLM cohorts application; and 3) The application of SFT and RLHF.

The application of large language models (LLMs) in the Internet of Things (IoT) and embedded systems has gained significant attention in recent years. Qiu et al.\cite{qiu2022edgeformer} proposed EdgeFormer, an edge-based transformer model for on-device natural language processing tasks in IoT environments. Their work demonstrated the feasibility of deploying LLMs on resource-constrained edge devices. Similarly, Zhang et al.\cite{zhang2022deflating} introduced a deflating technique to compress pre-trained LLMs for efficient deployment on embedded systems while maintaining performance.

Several works have explored the use of smaller-scale LLM cohorts for specific tasks. Su et al.\cite{su2022globalpipeline} proposed GlobalPipeline, a framework that decomposes large LLMs into smaller experts and orchestrates their collaboration. Their approach showed improved efficiency and scalability compared to monolithic LLMs. Likewise, Dai et al.\cite{dai2022knowledge} introduced a knowledge distillation method to train smaller LLMs from larger ones, enabling efficient deployment on edge devices.

Fine-tuning pre-trained LLMs has proven effective for adapting them to specific tasks and domains. Supervised fine-tuning (SFT) has been widely used to fine-tune LLMs on labeled data~\cite{gao2021making,lee2022deduplicating}. Reinforcement learning from human feedback (RLHF) has also been explored as a fine-tuning approach, where human feedback is used to refine the LLM's behavior~\cite{stiennon2020learning,ouyang2022training}.

\section{Threat Model}\label{section:threat-model}
In this section, we describe the potential attack vectors that can be detected by the LLM-empowered sub-system and the computing requirements for deploying LLMs at various levels of the system architecture.

Aero-LLM framework aims to detect and forecast the following attack vectors: 1) \textit{network attacks} that attempt to disrupt these channels through jamming, spoofing, or man-in-the-middle attacks; 2) \textit{sensor manipulation attacks} that manipulate sensor data or inject false information to mislead the UAV's decision-making processes; 3) \textit{software vulnerabilities} that can be exploited by adversaries to gain unauthorized access or control; and 4) \textit{insider threats} that attempt to disrupt operations or steal sensitive data.

A reliable and high-bandwidth network infrastructure is essential to enable real-time communication and data exchange between the UAV, edge servers, and cloud servers. For smaller LLMs deployed onboard the UAV, embedded GPUs or specialized AI accelerators with limited computational resources may be sufficient. At the edge and cloud level, more powerful GPU resources may be available, enabling the deployment of larger-scale LLMs for computationally intensive tasks, such as finetuning or inference with large context windows. By deploying LLMs at various levels of the system architecture, the Aero-LLM framework provides a robust and effective defense against various cyber threats targeting UAV systems.

\section{System Design}\label{section:system-design}

Areo-LLM uses a multi-tiered architecture with specialized LLMs deployed across onboard, edge, and cloud environments as follow:
1) \textit{On-board}: Small models like OPT-350m and OPT-125m are used for real-time tasks such as anomaly detection and basic inferencing, chosen for their low computational requirements and minimal latency. 2) \textit{At the edge}: Medium models, such as OPT-1.3B and OPT-6.7B, are deployed on edge servers with powerful GPUs for complex tasks like aggregated data analysis and intermediate anomaly detection, balancing computational power and latency. 3) \textit{In the cloud}: Large models including Llama2-7B, Llama2-13B, TimesNet, and Time-LLM are deployed in the cloud for comprehensive data analysis and long-term forecasting, utilizing extensive computational resources. This setup suits non-time-sensitive tasks despite higher latency. The rationale of this design is that onboard LLMs offer lowest latency with limited power. Edge LLMs balance the latency and computational capacity, while cloud LLMs provide high computing power but suffer from the increased latency. This hierarchical design balances performance and optimizes resource utilization.

\begin{figure*}[ht!]
    \centering
    \includegraphics[width=.6\textwidth]{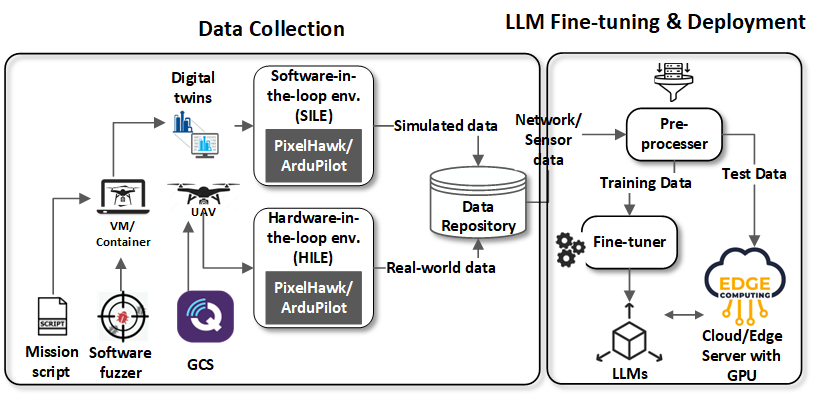}
    \caption{System Architecture of Aero-LLM}
    \label{fig:aero-gpt-system}
\end{figure*}

\section{Implementation}
Figure~\ref{fig:aero-gpt-system} illustrates the sequential phases of data collection, LLM fine-tuning, and deployment for UAV systems. The UAV data was collected from two sources: the digital twins and the actual UAVs connected with software-in-the-loop (SITL) and hardware-in-the-loop (HITL) environments, using PixHawk\cite{PixHawk} and ArduPilot\cite{ArduPilot} to simulate and capture data, which is then fed into a central repository. This repository collects both simulated and real-world sensing data, vital for fine-tuning task-specific LLMs. The fine-tuning of LLMs occurs on the cloud server, using the training data to ensure the models are well-adapted to the operational context of UAVs. Once fine-tuned, these models are deployed on edge computing platforms, leveraging the edge server equipped with GPUs. This enables efficient and rapid processing of UAV data in real-time, optimizing performance and responsiveness for mission-critical applications.

\subsection{LLM Fine-tuning and Deployment}\label{section:system-design:finetuning}

\begin{figure}[ht!]
    \centering
  \includegraphics[scale=.8]{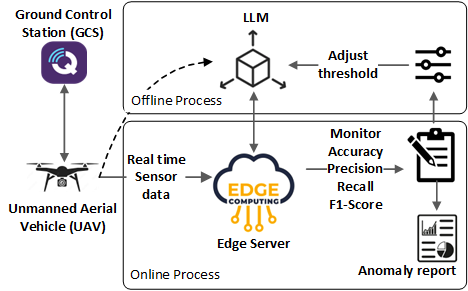}
  \caption{LLM Fine-tuning and Deployment}
    \label{fig:realtime-detection}
\end{figure}

Figure~\ref{fig:realtime-detection} illustrates the interaction between different components, including the GCS, the UAV, the cloud, and edge servers.
The cloud server performs offline processes, including fine-tuning the LLM for enhanced inferencing, ensuring the LLM’s responsiveness is precisely calibrated for anomaly detection.

During active missions, the UAV transmits real-time sensor data to the edge server for immediate processing—an online process leveraging edge computing to minimize latency. The edge Server employs the LLM to analyze incoming data streams in real time, ensuring swift anomaly detection. Critical performance metrics such as accuracy, precision, recall, and F1-Score are continuously monitored to maintain operational integrity.

When anomalies are detected, the system generates detailed anomaly reports. These reports serve as actionable insights for the GCS, informing potential mission adjustments and contributing to the iterative enhancement of the UAV's operational framework. This closed-loop system exemplifies the integration of edge AI and real-time analytics in UAV operations, prioritizing immediate data processing and adaptive learning for continual improvement of mission-critical tasks.

\section{Implementation}

\subsection{Preprocessing Data}
\subsubsection{Network layer data}

\begin{table*}[ht!]
  \centering
  \caption{The Format of Sensor Data}
  \label{table:sensor-data}
  \resizebox{\textwidth}{!}{ 
    \begin{tabular}{ |p{1cm}|p{1.6cm}|p{1.6cm}|p{1.6cm}|p{0.5cm}|p{0.3cm}|p{1.6cm}|p{1.6cm}|p{1.6cm}|p{1.6cm}|p{1.6cm}|  }
      \hline
      \multicolumn{1}{|c|}{Timestamp} & \multicolumn{1}{|c|}{Gyro 0 (rad/s)} & \multicolumn{1}{|c|}{Gyro 1 (rad/s)} & \multicolumn{1}{|c|}{Gyro 2 (rad/s)} & \multicolumn{1}{|c|}{Gyro Integral (dt)} & \multicolumn{1}{|c|}{Acc. Timestamp Relative} & \multicolumn{1}{|c|}{Acc. 0 (m/s²)} & \multicolumn{1}{|c|}{Acc. 1 (m/s²)} & \multicolumn{1}{|c|}{Acc. 2 (m/s²)} & \multicolumn{1}{|c|}{Acc. Integral (dt)} & \multicolumn{1}{|c|}{Acc. Clipping} \\
      \hline
      212000 & 0.004793696 & -0.002396833 & -0.000266331 & 4000 & 0 & 0.038307235 & 0.06703783 & -9.804258 & 4000 & 0 \\
      \hline
      220000 & 0.005592621 & -0.01171792 & 0.001864209 & 4000 & 0 & 0.001197101 & -0.033221554 & -9.8281975 & 4000 & 0 \\
      \hline
      224000 & 5.56E-09 & -0.005952639 & -0.000532643 & 4000 & 0 & 0.001197101 & -0.022744747 & -9.861719 & 4000 & 0 \\
      \hline
      228000 & -0.002130517 & 0.004793691 & -0.006125276 & 4000 & 0 & -0.005985504 & 0.013168283 & -9.858126 & 4000 & 0 \\
      \hline
      232000 & 0.00133158 & 0.004793683 & -0.006391593 & 4000 & 0 & -0.025139121 & -0.010773737 & -9.852142 & 4000 & 0 \\
      \hline
      240000 & 0.00378243 & 0.004216063 & -0.00506 & 4000 & 0 & 0.04908113 & -0.00119693 & -9.831791 & 4000 & 0 \\
      \hline
      244000 & -0.000798938 & 0.008788439 & -0.002929465 & 4000 & 0 & 0.053869545 & -0.021547647 & -9.837774 & 4000 & 0 \\
      \hline
      252000 & 0.001331579 & 0.009853696 & -0.003195797 & 4000 & 0 & 0.051475342 & 0.029927697 & -9.831791 & 4000 & 0 \\
      \hline
      256000 & 0.000798977 & 0.011984224 & 0.004793693 & 4000 & 0 & 0.028730419 & 0.02633639 & -9.805454 & 4000 & 0 \\
      \hline
    \end{tabular}
  }
\end{table*}

\begin{figure}
    \centering
  \includegraphics[scale=.5]{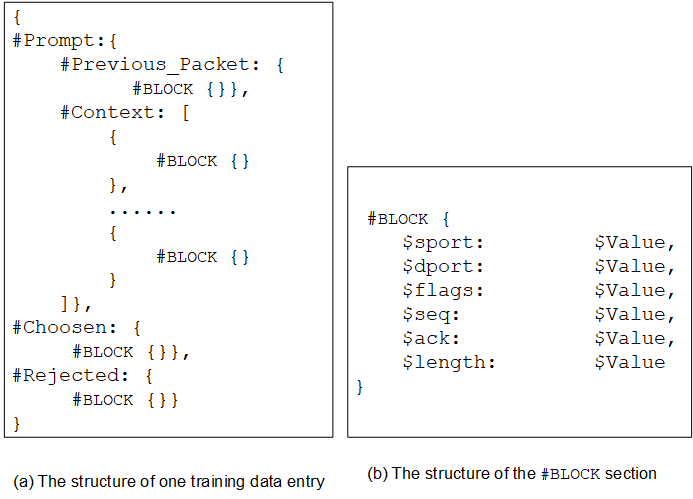}
  \caption{The Format of Data Under the fine-tuning.}
    \label{fig:training_data_format}
\end{figure}

We take the following steps to process network packets. First, we single out TCP sessions from our System-in-the-Loop (SIL) simulation. Then, we convert the sessions into a sliding window dataset. Using the format shown in Figure~\ref{fig:training_data_format}, we use one message as the prompt, along with previous $n$ messages serving as the context. Following this, we have packets labeled as '\textit{chosen}' and '\textit{rejected}.' The 'chosen' packet is the ground truth, representing the valid data in following packets of that session, providing positive feedback. The 'rejected' packet is the perturbation of one of the six key-value pairs, providing negative feedback. We apply the similar process for the collected TCP and UDP data, and MavLink sensor data for normal and anomalous missions using the Software in the Loop (SIL) simulation.

We compose the fine-tuning data in the format as illustrated in Fig.~\ref{fig:training_data_format}. Each data entry includes three sections: \texttt{\#Previous\_Packet}, \texttt{\#Predicted\_Packet}, and \texttt{\#Context}, which use the same \texttt{\texttt{\#BLOCK}} section that specifies  $<$\texttt{key:value}$>$ pairs. We describe how to process various data in following subsections.

\subsubsection{Application layer data}
We collect the sensor data from uLog generated from the missions as described in Section~\ref{section:system-design:mavlink}. The uLog data is parsed into CSV file using PX4 log utils. The CSV files are used for fine-tuning the models Time-LLM\cite{jin2023time} and TimesNet\cite{wu2023timesnet}. We chose combined sensor data for the experiments that contain sensor data from the Gyro and Accelerometer. The format of the data is shown in Table \ref{table:sensor-data}.

\subsection{Fine-tuning LLMs}\label{section:system-design:fine-tuning} 
\subsubsection{TCP/UDP Data}
The fine-tuning process utilizes a three-stage approach introduced by Microsoft's DeepSpeed~\cite{DeepSpeed}. The initial stage involves training an actor model through Supervised Fine-Tuning (SFT), a method where the model learns from a labeled dataset to predict accurate outputs. In the second stage, a critic model is trained through a human feedback loop, enabling the model to judge the accuracy of each value in a predicted packet with precision. The final stage, and arguably the most critical, involves Iterative Reinforcement Learning from Human Feedback (RLHF) Refinement. This stage deploys the RLHF model to further fine-tune the SFT model, resulting in a significantly enhanced final model that benefits from precise supervision and iterative feedback-driven adjustments.
\begin{table}[ht]
\centering
\caption{Accuracy of Predicted Fields (in \%)}
\label{tab:prediction_accuracy}
\resizebox{\columnwidth}{!}{ 
\begin{tabular}{lcc}
\hline
\textbf{Field} & \textbf{SFT \& RLHF }  \\
\hline
\textbf{OPT-350M, OPT-1.3B}   \\
\hline
sport & 100.00, 100.00\\
dport & 100.00, 100.00\\
flags & 99.85, 99.45\\
seq & 98.45, 98.34 \\
ack & 48.10, 53.97 \\
length & 99.95, 99.63 \\
\hline
For OPT-350M, 0 errors: 53.23, 1 error: 45.29,&2 errors: 1.11; 3 errors: 0.37, 4+ errors: 0\\
For OPT-1.3B, 0 errors: 47.30, 1 error: 51.80,&2 errors: 0.85; 3 errors: 0.05, 4+ errors: 0\\
\hline
\end{tabular}
}
\end{table}

\subsubsection{MavLink Sensor data}\label{section:system-design:mavlink}
PX4 sensor data is a time series. Various sensors in UAVs communicate by sending messages on various topics. We collect data from the Ulog and convert it into CSV format, which will be input into the model for fine-tuning.

We select two models for different tasks: TimesNet for anomaly detection and Time-LLM for forecasting. We first gather historical data from UAV sensors to fine-tune the Time-LLM for forecasting. This data undergoes preprocessing, including filling in missing values, normalization, and division into training, validation, and test subsets. It's then tokenized to match Time-LLM's language, trained on the training set, and evaluated on the test set to gauge its forecasting performance. Accuracy and precision are measured by calculating MAE and MSE. Finally, the optimized model is saved for future predictions.

To finetune the TimesNet model for anomaly detection, we commence by collecting UAV sensor data, such as altitude and GPS coordinates, followed by preprocessing that includes imputation of missing values and feature scaling. We divide the dataset into subsets, ensuring a mix of normal and anomalous points, and label them based on historical anomaly records. With the TimesNet model initialized with pre-trained weights, we then set the fine-tuning parameters, such as sequence length and learning rate. After training on the dataset until performance plateaus, the refined model is saved and ready for real-time anomaly detection in UAV operations.

\subsection{Cyber Security tasks}
We choose the TimesNet\cite{wu2023timesnet} model for anomaly detection and train it using the preprocessed PX4 Sensor data.After the model is fine-tuned using PX4 Sensor data, the anomaly detection script is run using the test data set with labels. 
  \begin{table}[ht]
  \centering
    \caption{Parameters for finetuning Large models} 
\label{table:timesnet-forecaster}
 \begin{tabular}{ |p{3cm}|p{2.5cm}|p{2.5cm}| }
 \hline
 \multicolumn{1}{|c|}{Training parameters} & \multicolumn{2}{|c|}{Model}\\
 \hline
 Parameter & TimesNet & Times-LLM \\
 \hline
 Epochs & 3 & 1\\
 \hline
 Learning Rate & 0.02 & 0.02\\
 \hline
 Llama layers & N/A & 8 \\
 \hline
 Batch Size & 128 & 4 \\
 \hline
 Model Dimension & 64 & 32 \\
 \hline
 FCN Dimension & 64 & 128\\
 \hline
 \end{tabular}
\end{table}

We choose Time-LLM~\cite{jin2023time} model based on the Llama-2-7B to finetune the model for forecasting. Time-LLM reprograms large language models for time series without altering the pre-trained foundation model.Time-LLM model performs better than state-of-the-art solutions in few-shot and zero-shot scenarios.

 We set the following parameters for fine-tuning TimesNet and Time-LLM models :\\

\alglanguage{pseudocode}
\begin{algorithm}[H]
\small
    \caption{Anomaly Detection using TimesNet}
    \label{algorithm:anomaly-detection}
    \begin{algorithmic}[1]
        \Inputs{$\mathcal{M}$: TimesNet Model, \par
                $\Theta$: Fine-tuned Model Weights and Biases, \par
                $\mathcal{D}$: Real-time Sensor Data, 
                $E$: Number of Epochs, \par
                $B$: Batch Size, 
                $A$: Anomaly ratio, 
                $L$: Training Loss }
        \Output{$\alpha$: Accuracy, $\pi$: Precision, $\rho$: Recall, $\phi$: F-Score}
        \Initialize{
            $\mathcal{P} \gets \emptyset$; $\mathcal{G} \gets \emptyset$\par
            $\tau \gets predefined\_value$; $e \gets 1$
        }
        \State $ \mathcal{M} \gets load\_weight\_bias(\Theta) $
        \While{$e \leq E$}
            \ForEach{$batch \in \mathcal{D}$}
                \State $x, y \gets batch$
                \State $p \gets \mathcal{M}.predict(x)$
                \State Append $p$ to $\mathcal{P}$; Append $y$ to $\mathcal{G}$
            \EndFor
            \State $e \gets e + 1$
        \EndWhile
        \State $threshold \gets percentile(L, 100 - A)$
        \State $loss \gets MSE(\mathcal{P} - \mathcal{G})$  
        \State $ \mathcal{P} \gets loss > threshold $
        \State Calculate evaluation metrics:
        \State $TP \gets  count(\mathcal{P} = 1 \wedge \mathcal{G} = 1);
         TN \gets  count(\mathcal{P} = 0 \wedge \mathcal{G} = 0)$
        \State $FP \gets  count(\mathcal{P} = 1 \wedge \mathcal{G} \neq 1);
        FN \gets  count(\mathcal{P} = 0 \wedge \mathcal{G} \neq 0)$
        \State $\alpha \gets \frac {TP}{TP + FP};
        \pi \gets \frac{TP}{TP + FP};\rho \gets \frac{TP}{TP + FN}$ 
        \State $\phi \gets weighted\_harmonic\_mean(\pi, \rho)$ \Comment{F-Score}
        \State \textbf{Return} $\alpha, \pi, \rho, \phi$
    \end{algorithmic}
\end{algorithm}
The metrics from fine-tuning: 
  \begin{table}[ht!]
  \centering
    \caption{Metrics From TimesNet Finetuning} 
\label{table:timesnet-finetuning-metrics}
\begin{tabular}{ |p{3cm}|p{3cm}|  }
\hline
\multicolumn{2}{|c|}{Metrics} \\
\hline 
Train Loss & 0.1713483 \\
\hline 
Validation Loss & 0.1251146 \\
\hline 
Test Loss & 0.1068001 \\
\hline 
MAE Loss & 0.2587003 \\
\hline
Learn Rate & 0.0008000000 \\
\hline
\end{tabular}
\end{table}

\section{Evaluation}\label{section:evaluation}
\subsubsection{Anomaly Detection}
To enrich our dataset with anomalous data, we employ several techniques to transform normal records. First, we designate every n-th record as anomalous. Additionally, we introduce irregularities by randomly altering select records. To simulate different degrees of deviation, we systematically vary the dataset's variance and generate corresponding datasets. Lastly, we apply a Poisson distribution to intersperse anomalous data throughout the dataset, ensuring a realistic distribution of anomalies for robust model training.

The following formulas are used for detecting anomalies. 
The loss is calculated as:
\begin{equation}\label{loss_formula}
 loss = MSE(predicted\_value - groundtruth\_value)
\end{equation}
The anomaly threshold is calculated as:
\begin{equation}\label{anomaly_threshold}
 threshold = Percentile(loss, 100 - anomaly\_ratio)
\end{equation}

\paragraph{Every n\textsuperscript{th} record is anomaly}
We take the normal data from the sensor and manipulate every n\textsuperscript{th} record to contain anomalous data. For the experiment, we choose n=5. The anomalous data thus generated is fed into the TimesNet model running on the Edge Server, and the loss is calculated as per formula \ref{loss_formula}. When the loss exceeds the thresholds as calculated in formula \ref{anomaly_threshold}, the data is flagged as an anomaly. The anomaly metrics are illustrated in Figure~\ref{fig:timesnet-anomaly-det}

\begin{figure}[ht!]
    \centering
  \includegraphics[scale=.3]{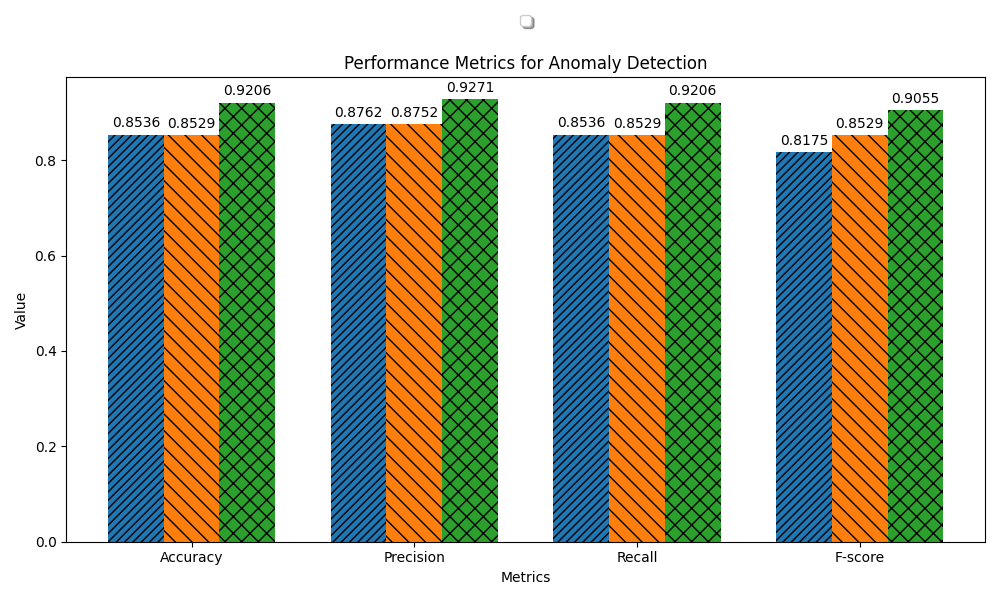}
  \caption{Metrics From TimesNet Anomaly Detection}
    \label{fig:timesnet-anomaly-det}
\end{figure}

\begin{figure}[ht!]
    \centering
    \includegraphics[scale=.2]{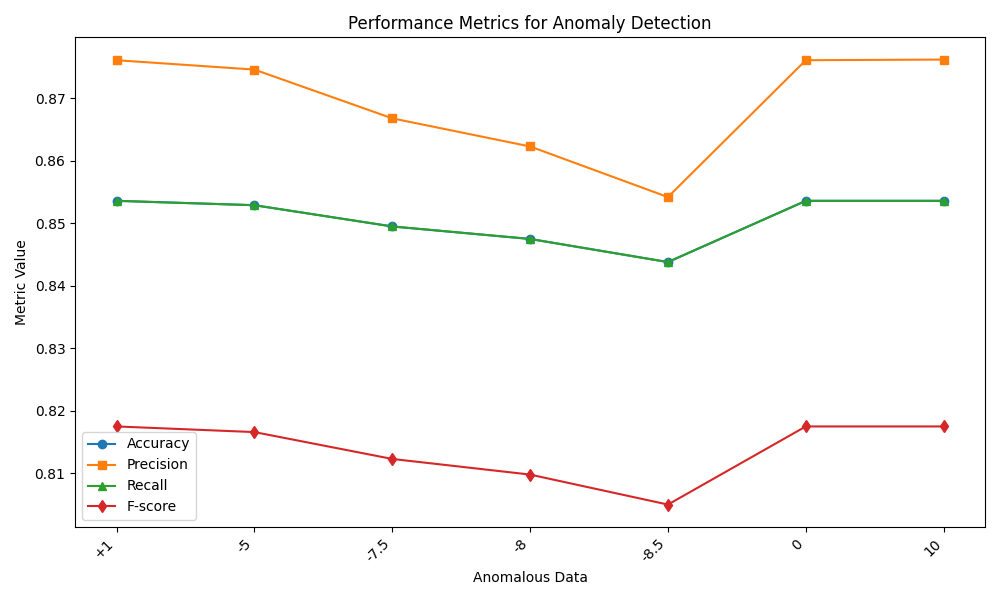}
    \caption{Anomalous Data vs Metrics}
    \label{fig:timesnet-anomaly-det}
\end{figure}

\paragraph{Change variance of anomalous data}
Sensor reading accelerometer\_m\_s2\_2 is arbitrarily changed to simulate variances and the anomaly detection script is run for each variance and the data is collected. The metrics from changing the variance is calculated and plotted in Figure The anomaly metrics are illustrated in Figure~\ref{fig:timesnet-anomaly-det}.

\begin{figure}[ht!]
    \centering
    \begin{subfigure}{0.49\columnwidth}
        \includegraphics[width=\linewidth]{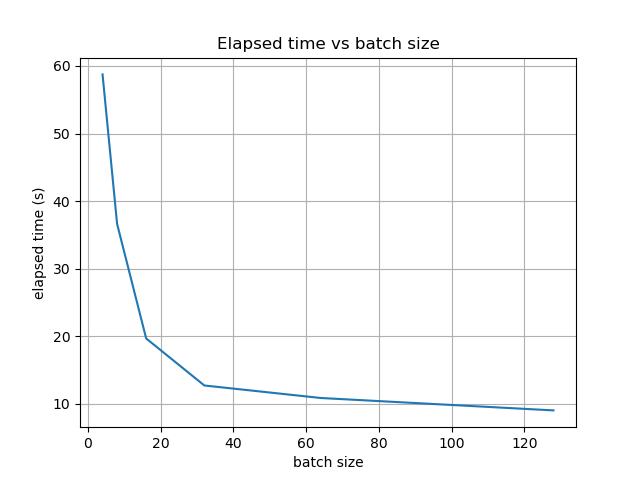}
        \caption{Batch vs elapsed time}
        \label{fig:batch_vs_time}
    \end{subfigure}
    \begin{subfigure}{0.49\columnwidth}
        \includegraphics[width=\linewidth]{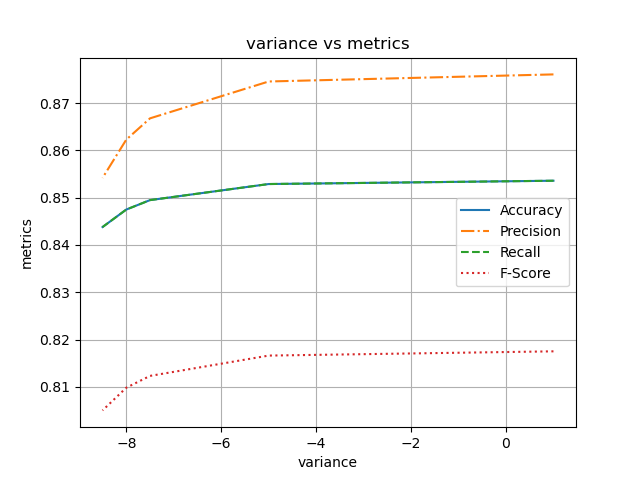}
        \caption{Variance vs Metrics}
        \label{fig:variance_metrics}
    \end{subfigure}
    \caption{Comparative analysis of TimesNet performance}
    \label{fig:timesnet_performance}
\end{figure}

From the test results and the graph in figure \ref{fig:variance_metrics}, we can infer that the TimesNet model is very resilient, and the variance in data doesn't seem to impact a lot on the metrics such as accuracy, precision, recall, and F-Score.

\paragraph{Impact of batch size} 
We evaluate the relationship between the batch size and time to detect anomalies. We set one sensor reading to an arbitrary value of -8.5 and run the anomaly detection script by just changing the batch size and keeping all other parameters. All the runs produce the same metrics as
Accuracy : 0.8438, Precision : 0.8542, Recall : 0.8438, F-score : 0.8050.
\begin{table}[ht!]
  \centering
    \caption{Metrics From TimesNet for various batch size } 
\label{table:timesnet-batch-anomaly-det}
\begin{tabular}{ |p{3cm}|p{3cm}|  }
\hline
\multicolumn{2}{|c|}{Metrics - Batch size} \\
\hline 
Batch Size & Avg Elapsed Time\\
\hline
128 & 9.08162 \\
\hline
64 & 10.89548 \\
\hline
32 & 12.71874 \\
\hline
16 & 18.21896 \\
\hline
8 & 34.2963 \\
\hline
4 & 61.91978 \\
\hline
\end{tabular}
\end{table}
From the table \ref{table:timesnet-batch-anomaly-det} of experiment results and from the graph \ref{fig:batch_vs_time}, we can infer that the batch size plays a significant role in the sensitivity of the anomaly detection. Batch size 48 and above significantly reduced the elapsed time to detect anomalies. Also, increasing the batch size beyond 48 doesn't reduce the elapsed time proportionally.

Please note that the accuracy and recall lines are not distinguishable as the accuracy and recall data are the same. So, the lines appear in Green color.
\subsubsection{Forecasting}
We run the forecasting script that uses a finetuned Time-LLM model for the test dataset, and We record the loss for the test data. The test data is taken from one mission and consists of normal data. The model is able to predict the future data accurately. We calculate the test Loss is equal to  \textit{0.1068001} and MAE Loss is equal \textit{0.2587003}

The results show the accuracy is > 82\%.

\section{Conclusion}\label{section:conclusion}

In this paper, we present the Aero-LLM framework for UAVs , addressing the need for secure, efficient, and intelligent systems. Aero-LLM integrates Large Language Models (LLMs) to enhance UAV capabilities for complex missions, ensuring security and efficiency. Future research aims to extend Aero-LLM to support multiple UAVs with individual computational LLMs for autonomous decision-making, enabling collaboration in terrain mapping and mission execution. Integrating advanced sensors and real-time data processing will enhance operational accuracy and efficiency. Aero-LLM seeks to achieve higher mission success and reliability through a distributed network of intelligent UAVs. Comprehensive evaluation confirms Aero-LLM's high performance across key metrics such as accuracy, precision, recall, and F1 score, while maintaining a low memory footprint.

\section{Acknowledgment}
This work was supported in part by the National Aeronautics and Space Administration (NASA) award 80NSSC20M0124 and Oakland University URC faculty research fellowship. 

\bibliographystyle{IEEEtran}
\bibliography{ICCCN_2024_05202024_Final}

\end{document}